\documentclass{PoS}

\newcommand{\be}{\begin{equation}}
\newcommand{\ee}{\end{equation}}

\def\ff{f\hspace{-0.3cm}f}

\title{NNLL resummation for the associated production of a top pair with a heavy boson at the LHC}

\ShortTitle{NNLL resummation for the associated production of a top pair with a heavy boson at the LHC}

\author{\speaker{Alessandro Broggio}%
	    \thanks{Preprint number: TUM-HEP-1125/18}\\
        Physik Department T31, Technische Universit\"at M\"unchen,
        James Franck-Stra{\ss}e 1, D-85748 Garching, Germany\\
        E-mail: \email{alessandro.broggio@tum.de}}

\abstract{In this presentation we review recent results on the resummation of soft gluon emission corrections for the associated production of a top-quark pair with a heavy boson (Higgs/W/Z) at the Large Hadron Collider (LHC). We develop a parton level Monte Carlo based on a soft-gluon resummation formula valid up to next-to-next-to-leading logarithmic (NNLL) accuracy.
With this tool we study the impact of the newly computed corrections to the total cross sections and some important differential distributions.}

\FullConference{13th International Symposium on Radiative Corrections (Applications of Quantum Field Theory to Phenomenology)\\
		24-29 September, 2017\\
		St. Gilgen, Austria}

\begin{document}

\section{Introduction}

The production process of a top-antitop pair in association with a Higgs boson provides direct information on the top-quark Yukawa coupling. Indeed the tree level cross section for this process is proportional to the square of this coupling. Its precise knowledge is fundamental for verifying the origin of fermion masses and to test the SM prediction.
For these reasons the measurement of this particular Higgs production mode by the experimental collaborations is one of the goals of the current run of the LHC. Very recently there has been evidence for the $t \bar{t} H$ production channel with a first determination of the total cross section by the ATLAS collaboration  $\sigma(t \bar{t} H)=590^{+160}_{-150}$fb \cite{ATLAS:2017lpi} for the LHC operating at the energy of 13 TeV. This value is currently in agreement with the SM predictions. Eventual deviations of the measured cross section from the predicted SM value could be a hint of new physics. Due to the importance of this process, a long list of calculations has been performed in the past years to improve its theoretical accuracy.
The next-to-leading (NLO) QCD corrections were first evaluated in \cite{Beenakker:2002nc,Dawson:2003zu}. In more recent years NLO QCD corrections were computed using automated tools and interfaced with Monte Carlo event generators including parton shower effects \cite{Frederix:2011zi,Garzelli:2011vp,Hartanto:2015uka}.
Electroweak (EW) corrections for this process were computed in \cite{Yu:2014cka,Frixione:2015zaa}. NLO QCD and EW corrections considering the decay of the top quarks and off-shell effects were obtained in \cite{Denner:2015yca,Denner:2016wet}.
The resummation of higher order soft-gluon emission corrections were considered to next-to-leading logarithmic (NLL) accuracy in the production threshold limit in \cite{Kulesza:2015vda} and to next-to-next-to-leading logarithmic (NNLL) accuracy in the partonic threshold limit (or ``Triple Invariant Mass" kinematics limit) in \cite{Broggio:2015lya,Broggio:2016lfj,Kulesza:2017ukk}. Studies of the top-quark Yukawa coupling in the presence of a pseudoscalar component have been performed in \cite{Broggio:2017oyu} to NLO+NLL accuracy.

The cross sections for the associated production of a top pair with a Z or a W boson were measured at the LHC at 13 TeV both by the ATLAS \cite{Aad:2015eua} and CMS \cite{CMS:2017uib} collaborations.
The $t\bar{t}Z$ process is particularly important since it allows one to study the coupling of the $Z$ boson to the top quark. This measurement tests the SM predictions and eventually constraints beyond the SM scenarios which predict a deviation from the SM value. Both the $t\bar{t} W$ and $t\bar{t} Z$ processes have high multiplicity finale states and, for this reason, they are considered background processes in the search for heavy particles decaying via long chains, such as supersymmetric partners.
NLO QCD and EW corrections to the $t \bar{t} W$ and $t \bar{t} Z$ processes were computed by several groups in \cite{Lazopoulos:2007bv,Lazopoulos:2008de,Garzelli:2012bn,Campbell:2012dh,Maltoni:2014zpa,Maltoni:2015ena,Frixione:2015zaa,Frederix:2017wme,Rontsch:2014cca,Rontsch:2015una}. The resummation of soft-gluon emission corrections to NNLL accuracy matched to NLO calculations were computed in \cite{Li:2014ula,Broggio:2016zgg,Broggio:2017kzi}.

This talk is based on the following papers \cite{Broggio:2015lya,Broggio:2016lfj,Broggio:2016zgg,Broggio:2017kzi} and has the purpose to present the phenomenological impact of the resummed soft emission corrections to the total cross sections and differential distributions for the $t\bar{t} H$, $t\bar{t} W$, $t\bar{t} Z$ production processes at the LHC. We work in the soft-collinear effective theory (SCET) framework\footnote{See for example \cite{Becher:2014oda} for an introduction to the effective theory methods.} in Mellin space. We update some of the previously published results with the most recent experimental measurements.

\section{Factorization and resummation}

The associated production of a top pair and a Higgs or a $Z$ boson receives contributions at the lowest order in QCD from the partonic processes
\begin{displaymath}
i(p_1)+j(p_2) \rightarrow t(p_3) + \bar{t}(p_4) + H/Z(p_5) + X,
\end{displaymath}
where $i,j \in \{q\bar{q},\bar{q}q,gg\}$. $X$ indicates the unobserved partonic final-state radiation. In the case of the $t \bar{t} W$ process only the quark-initiated channel is present at the lowest order in QCD
\begin{displaymath}
i(p_1) +j(p_2)\rightarrow t(p_3) + \bar{t}(p_4) + W^{\pm}(p_5) + X,
\end{displaymath}
where $i,j\in q, \bar{q}^\prime$: $i$ represents a light up-type quark and $j$ a down-type light quark.
We define two Mandelstam invariants which are relevant for our discussion
\begin{displaymath}
\hat{s}=(p_1+p_2)^2 = 2p_1\cdot p_2 \, ,\quad \textrm{and}\quad M^2=(p_3+p_4+p_5)^2 .
\end{displaymath}
These two quantities coincide at LO, but when real radiation is emitted in the final state, one can take the ratio of the invariants $z\equiv M^2/\hat{s}$ and define the soft or partonic threshold region for $z\to 1$. We stress that in this limit the final state radiation $X$ can only be soft.
The factorization formula for the QCD cross sections in the partonic threshold limit is the same for all three processes and it was first derived in \cite{Broggio:2015lya} for the $t \bar{t} H$ case
\begin{displaymath}
\sigma \left(s,m_t,m_V \right) = \frac{1}{2 s} \int_{\tau_{\mbox{\footnotesize min}}}^{1} \!\!\! d \tau \int^1_{\tau} \frac{dz}{\sqrt{z}} \sum_{ij} \ff_{ij} \left( \frac{\tau}{z}, \mu \right) \int d\textrm{PS}_{t\bar{t}V} \mbox{Tr}\left[\mathbf{H}_{ij}\left(\{p\},\mu\right) \mathbf{S}_{ij}\left(\frac{M (1-z)}{\sqrt{z}},\{p\},\mu\right)  \right] \, ,
\label{eq:factorization}
\end{displaymath}
where $d\textrm{PS}_{t\bar{t}V}$ indicates the reduced treelevel 3-body phase space and $V=\{H,Z,W\}$. The symbols $\ff_{ij}$ represent the luminosity functions which depend on the partonic channel. The quantity $s$ is the square of the hadronic center-of-mass energy and we defined $\tau_{\mbox{\footnotesize min}}=(2 m_t+m_V)^2/s$ and $\tau=M^2/s$. The hard functions $\mathbf{H}_{ij}$, which are matrices in color space, are obtained from the color-decomposed one loop virtual corrections to the $2\to3$ tree-level processes. The soft functions $\mathbf{S}_{ij}$ represent the color-decomposed real emission corrections in the soft limit. They depend on plus distributions in the $z$ variable as well as on Dirac delta function of argument $(1-z)$. The precise form of the singular distributions can be found in \cite{Broggio:2015lya,Broggio:2016lfj}. The hard and the soft functions satisfy renormalization group equations (RGE) which are controlled by anomalous dimension matrices $\Gamma^{ij}_{H,S}$. In order to carry out the resummation of soft-gluon emission corrections to NNLL accuracy, the hard functions, soft functions and anomalous dimensions need to be computed up to NLO in $\alpha_s$. The NLO soft functions and anomalous dimensions were obtained in \cite{Li:2014ula,Broggio:2015lya,Broggio:2016lfj} and are the same for all of the three processes and they depend on the partonic channel. The hard functions are instead process dependent and receive contributions only from the one-loop virtual corrections. We evaluate these color-decomposed one-loop amplitudes by customizing the loop provider {\tt{Openloops}}  \cite{Cascioli:2011va} used together with the library {\tt{Collier}} {\cite{Denner:2016kdg}}. The NLO hard functions for all the three processes have been cross-checked numerically by means of a modified version of {\tt{Gosam}} \cite{Cullen:2014yla,Broggio:2017vlv}.

We evaluate the resummation formula in Mellin space by taking the Mellin transform of the cross section
\begin{displaymath}
\sigma(s,m_t,m_V) =  \frac{1}{2 s} \int_{\tau_{\textrm{min}}}^{1} \frac{d \tau}{\tau} \frac{1}{2 \pi i} \int_{c - i \infty}^{c + i \infty} dN \tau^{-N} \sum_{ij} \widetilde{\ff}_{ij}\left(N, \mu \right) \int d \textrm{PS}_{t \bar{t} V} \, \widetilde{c}_{ij} \left(N,\mu\right) \, ,
\label{eq:Mellinfac}
\end{displaymath}
where $\widetilde{\ff}_{ij}$ and $\widetilde{c}_{ij}$ are respectively the Mellin transforms of the luminosity functions and of the product of the hard and soft functions. See for exmaple \cite{Broggio:2016lfj,Broggio:2016zgg} for more details. The partonic threshold region $z\to 1$ corresponds to the limit $N\to \infty$ of the Mellin variable. The hard and soft functions can be evaluated in fixed order in perturbation theory at scales in which they are free from large logarithmic corrections. We indicate these scales with $\mu_h$ (hard scale) and $\mu_s$ (soft scale) respectively and we set their central values to $\mu_{h,\,0}=M$ and $\mu_{s,\, 0}=M/\bar{N}$, where $\bar{N} = N e^{\gamma_E}$. It is then possible to solve the RGEs for the hard and soft functions and evolve the hard-scattering kernel $\widetilde{c}_{ij}$ to the factorization scale $\mu_f$ which is the scale at which the parton densities are evaluated. The expressions for the resummed hard-scattering kernels in Mellin space assume the form
\begin{displaymath}
\widetilde{c}_{ij}(N,\mu_f) =  
\mbox{Tr} \Bigg[\widetilde{\mathbf{U}}_{ij}(\!\bar{N},\{p\},\mu_f,\mu_h,\mu_s) \, \mathbf{H}_{ij}( \{p\},\mu_h) \, \widetilde{\mathbf{U}}_{ij}^{\dagger}(\!\bar{N},\{p\},\mu_f,\mu_h,\mu_s) \widetilde{\mathbf{s}}_{ij}\left(\ln\frac{M^2}{\bar{N}^2 \mu_s^2},\{p\},\mu_s\right)\Bigg] \,  ,
\label{eq:Mellinresum}
\end{displaymath}
where the large logarithms of the ratio of the scales $\mu_h$ and $\mu_s$ are resummed by the evolution functions $\widetilde{\mathbf{U}}$, which are also matrices in color space. Their explicit expressions can be found in \cite{Broggio:2016lfj}.

\section{Numerical results}

The NNLL calculations are carried out by means of a in-house parton level Monte Carlo which is used to evaluate the resummed soft-gluon emission corrections. The NLO predictions are obtained  with \verb|MadGraph5_aMC@NLO| \cite{Alwall:2014hca}. We employ MMHT 2014 PDFs \cite{Harland-Lang:2014zoa} to the correspoding perturbative order of the calculation (for fixed-order predictions) and we use NNLO PDFs for NLO+NNLL results.
Our best prediction (NNLL) is matched to fixed order NLO calculations through the following matching formula to avoid double counting of contributions present in both computations
\begin{displaymath}
	\sigma^{{\textrm{NLO+NNLL}}}  =  \sigma^{{\textrm{NLO}}}
	+\left[ \sigma^{\textrm{NNLL}}- \sigma^{\textrm{approx. NLO}}\right]\,,
	\label{eq:NLOpNNLLmatching}
\end{displaymath}
where $\sigma^{\textrm{approx. NLO}}$ contains the $\mathcal{O}(\alpha_s)$ leading contributions in the soft emission limit.
We already discussed in the previous section the central scale choices for the hard and soft scales in the resummed calculations. In addition both fixed order and resummed computations depend on the choice of the factorization scale. We discussed in detail this scale choice in \cite{Broggio:2016lfj,Broggio:2016zgg} and we adopt the dynamical scale $\mu_f=M/2$. In fixed order calculations the uncertainty related to the particular choice of $\mu_f$ is estimated by varying this scale in the interval $\mu_f\in [\mu_{f,0}/2,2 \mu_{f,0}]$. Resummed results also depend on the scales $\mu_h,\mu_s$, hence the scale uncertainty is evaluated by varying  separately all three scales around their central values in the interval $\mu_i \in [\mu_{i,0}/2, 2\mu_{i,0}]$ for $i \in \{h,s,f\}$ and by finally combining them in quadrature (for details see \cite{Broggio:2017kzi}).

\begin{figure}[tp]
	\begin{center}
		\begin{tabular}{c}
			\includegraphics[width=10cm]{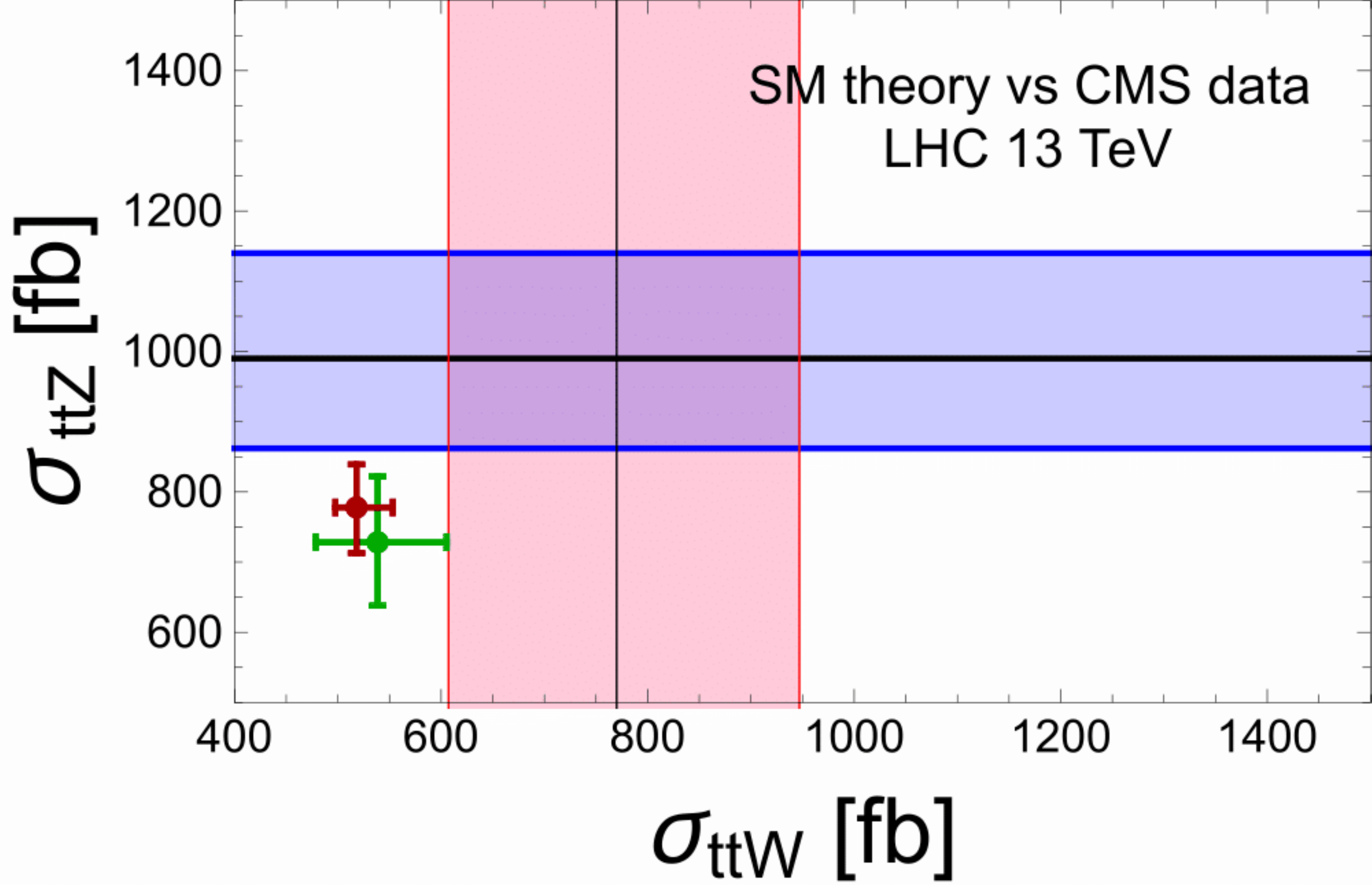}
		\end{tabular}
	\end{center}
	\caption{Total cross section at NLO (green cross) and NLO+NNLL (red cross) compared to the CMS measurements at $13$~TeV \cite{CMS:2017uib} (light blue and pink bands). \label{fig:totCS8and13}}
	
\end{figure}

Table \ref{tab:ttZttW} contains the total cross sections values for the three processes at NLO and NLO+NNLL accuracy. The latter ones are the  main results presented in this talk. By looking at the NLO+NNLL predictions for the $t \bar{t} H$ and $t \bar{t} Z$ processes we notice that, for this particular choice of the factorization scale, the central value of the cross sections is increased respect to the NLO calculations while for the $t \bar{t} W^{\pm}$ processes the central value is slightly decreased.
In all three cases the uncertainty bands at NLO+NNLL overlap nicely with the lower accuracy predictions and the central values fall within the NLO bands.
In Figure \ref{fig:totCS8and13} we graphically compare our theoretical calculations with the most recent CMS measurements \cite{CMS:2017uib} of the cross sections for the $t \bar{t} W$ and $t \bar{t} Z$ processes. The green cross corresponds to the NLO calculations while the red cross represents instead the NLO+NNLL calculations, both including scale(s) variation uncertainties.
The light red and light blue bands correspond respectively to the experimental determinations (including statistical and systematic uncertainties) of the $t \bar{t} W$ and $t \bar{t} Z$ cross sections. We find agreement between the theory predictions and the measurements considering that PDFs uncertainties are not taken into account in the comparisons.

\begin{figure}[tp]
	\begin{center}
		\begin{tabular}{cc}
			\includegraphics[width=7cm]{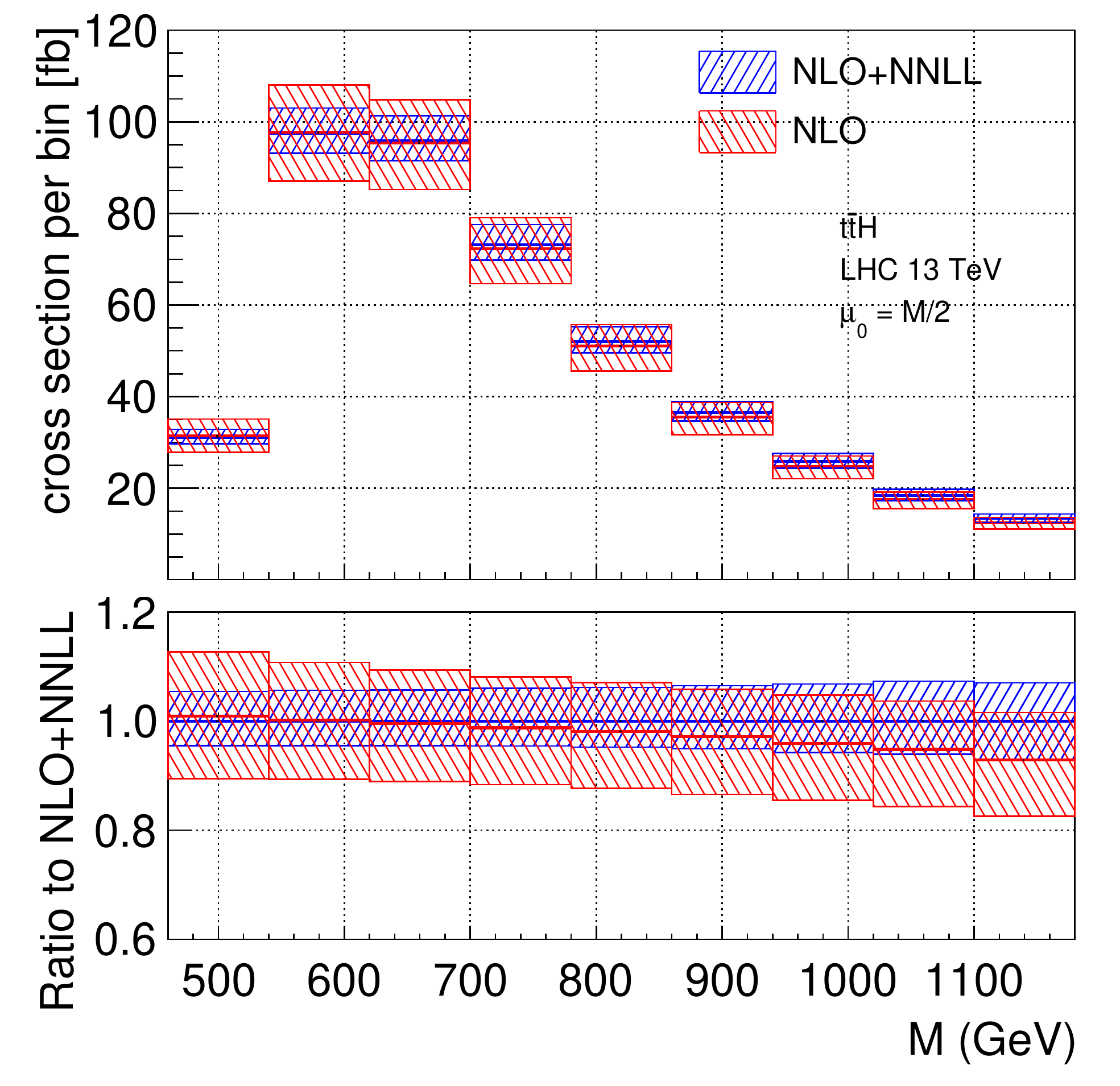} & \includegraphics[width=7cm]{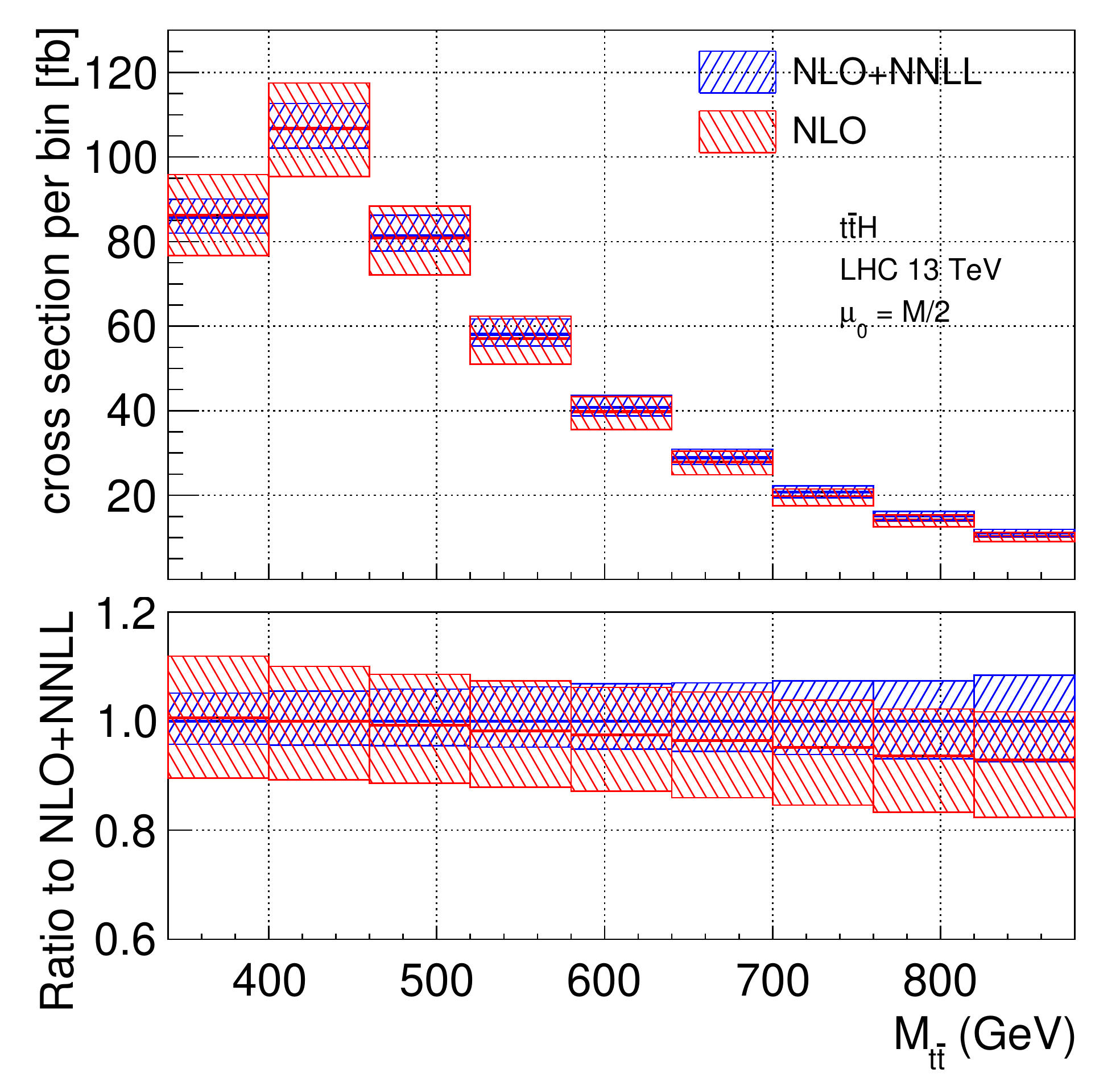} \\
			\includegraphics[width=7cm]{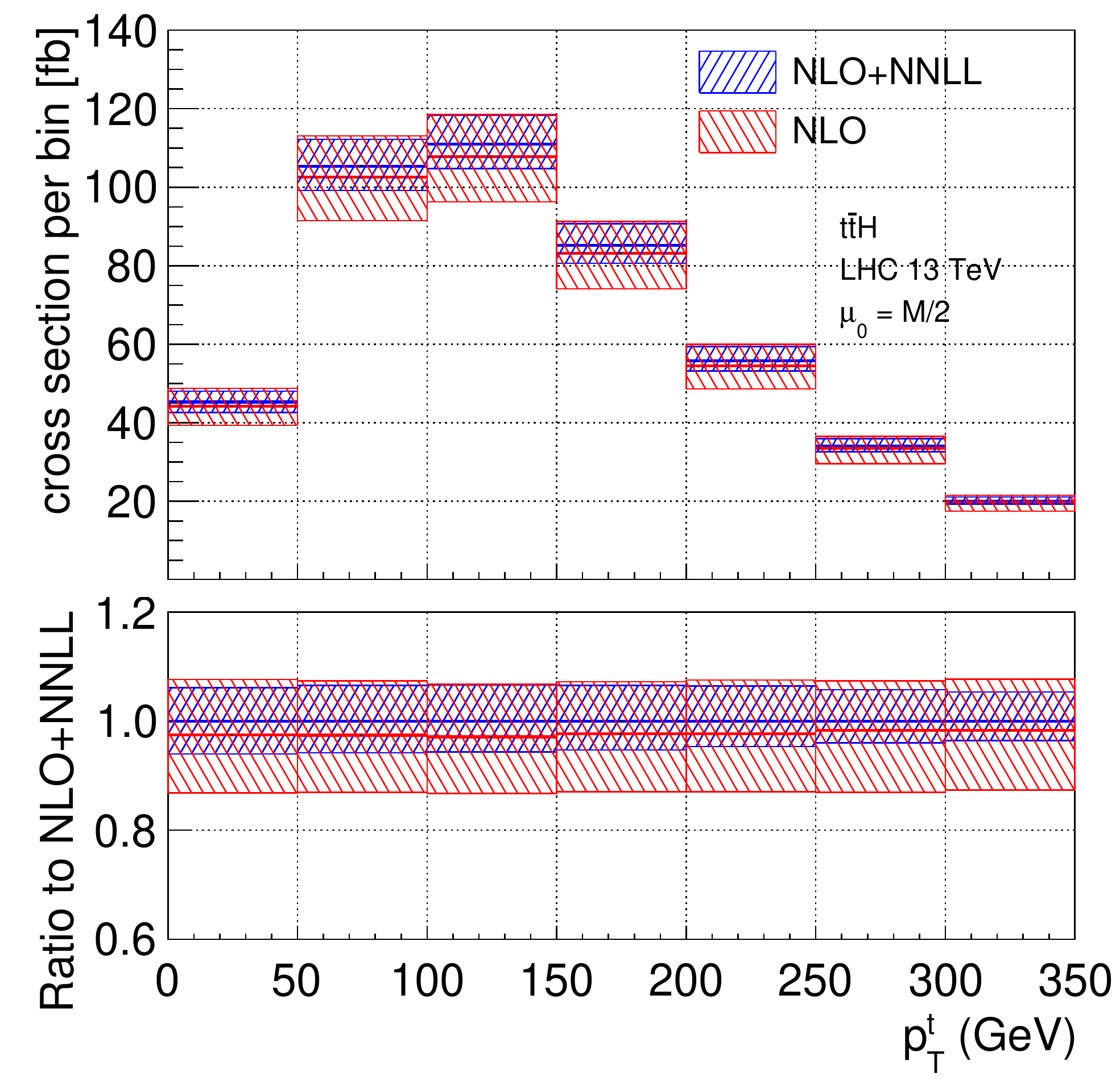} & \includegraphics[width=7cm]{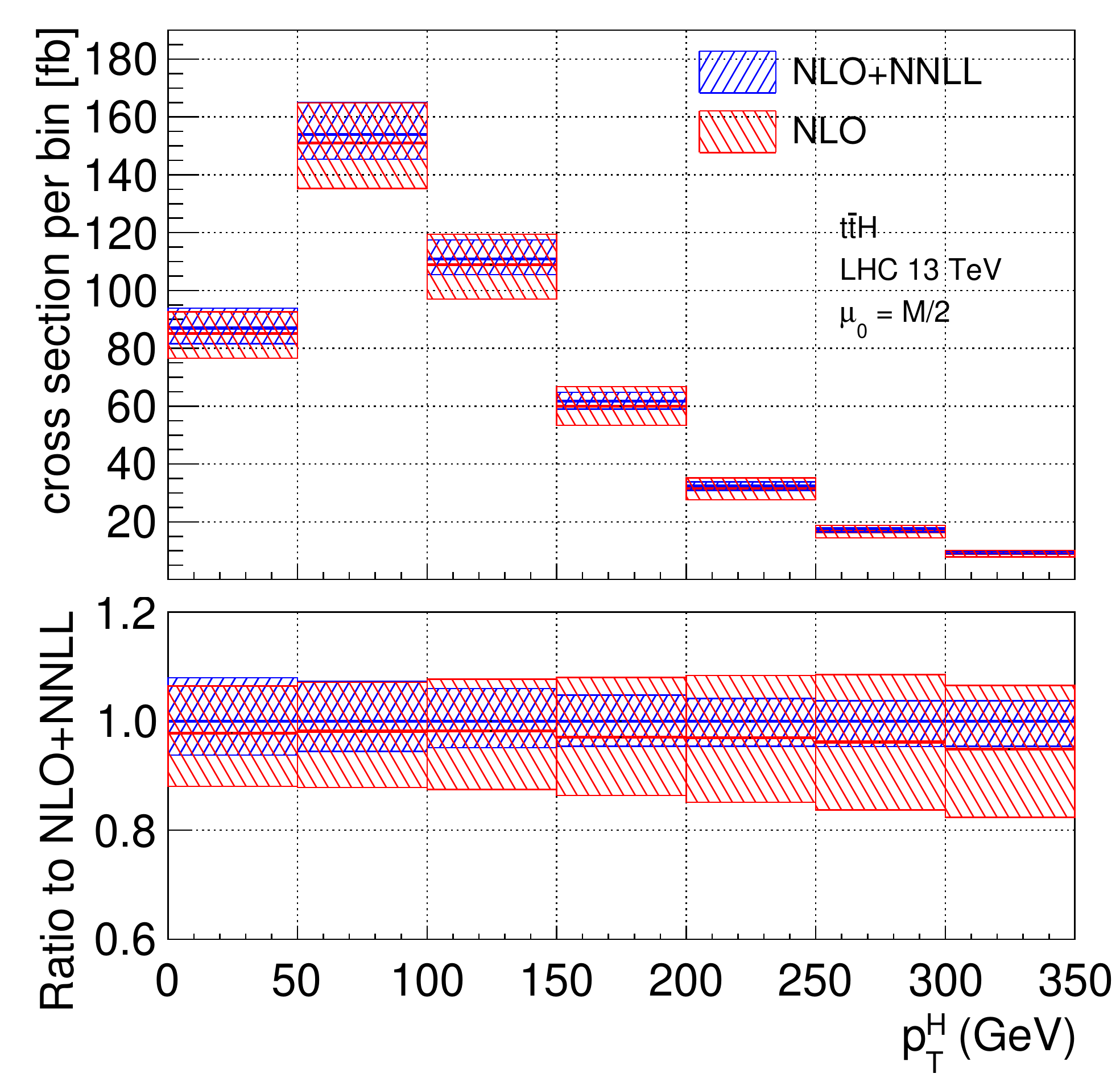} \\
		\end{tabular}
	\end{center}
	\caption{Differential distributions for $t \bar{t} H$ production at NLO+NNLL (blue band) compared to the NLO calculation (red band).
		The uncertainty bands are generated through scale variations of $\mu_f$, $\mu_s$ and $\mu_h$ as explained in \cite{Broggio:2016lfj}.
		\label{fig:NLOvsNNLLhalfMttH}
	}
\end{figure}

\begin{figure}[tp]
	\begin{center}
		\begin{tabular}{cc}
			\includegraphics[width=7cm]{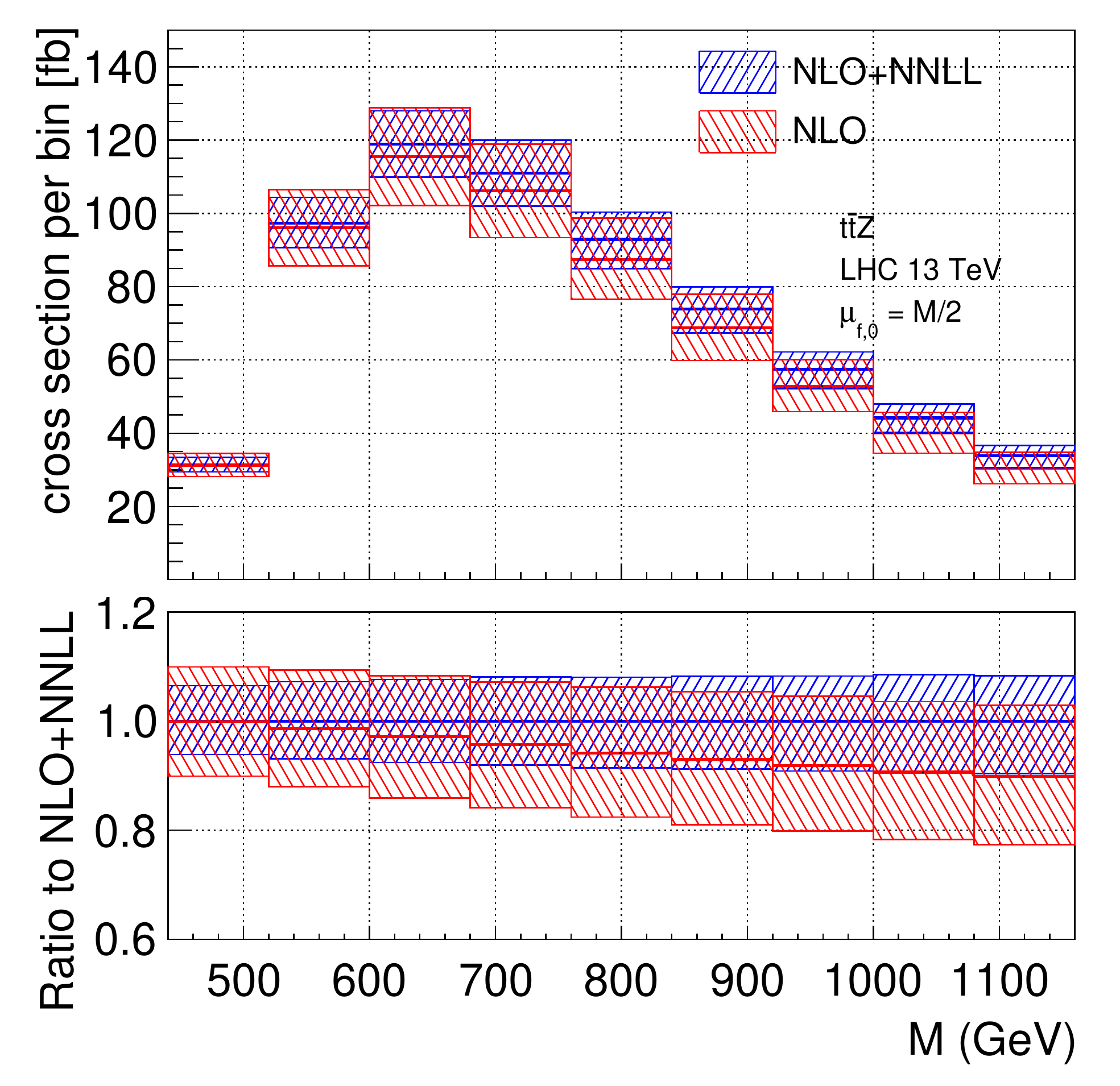} & \includegraphics[width=7cm]{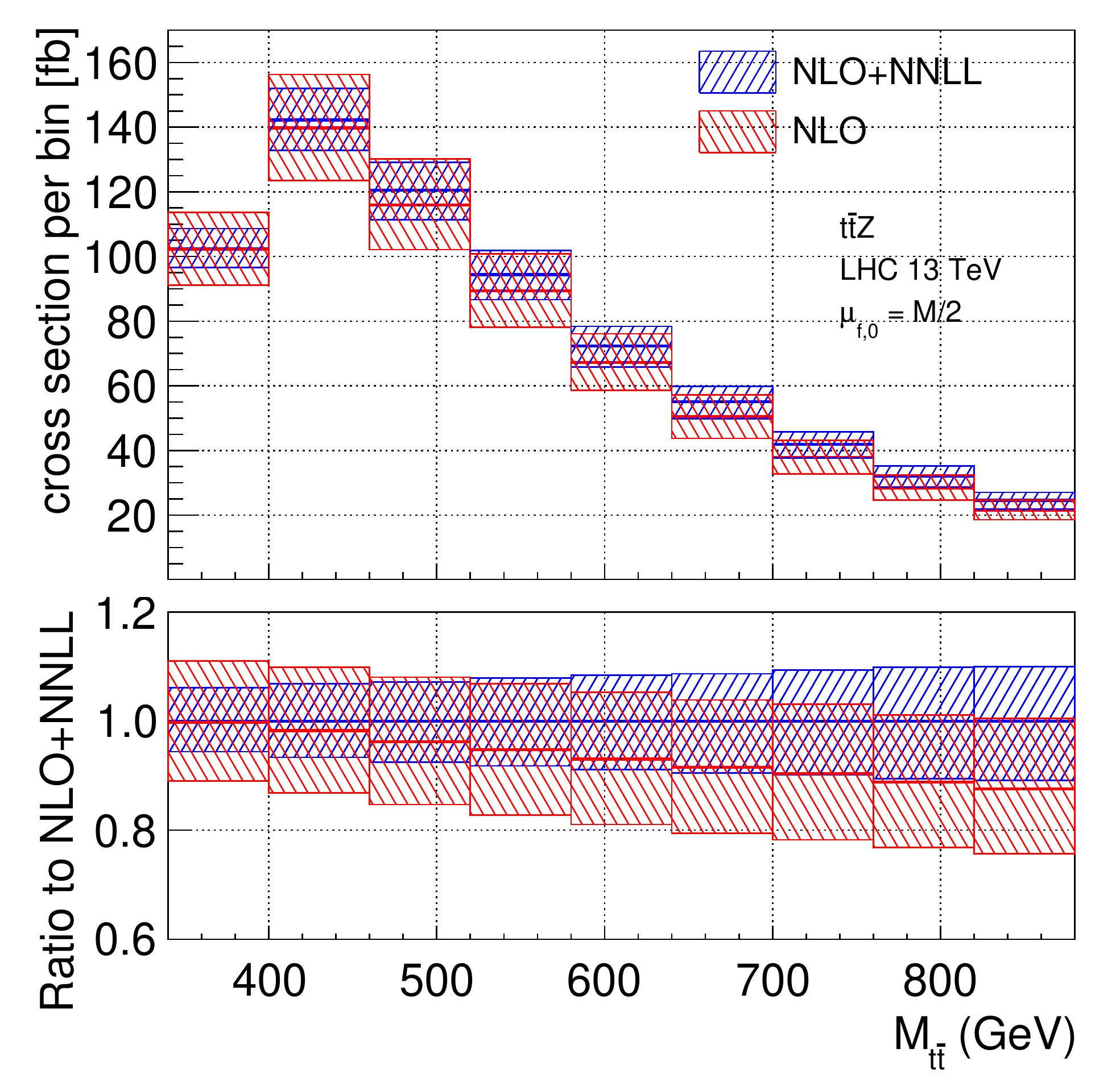} \\
			\includegraphics[width=7cm]{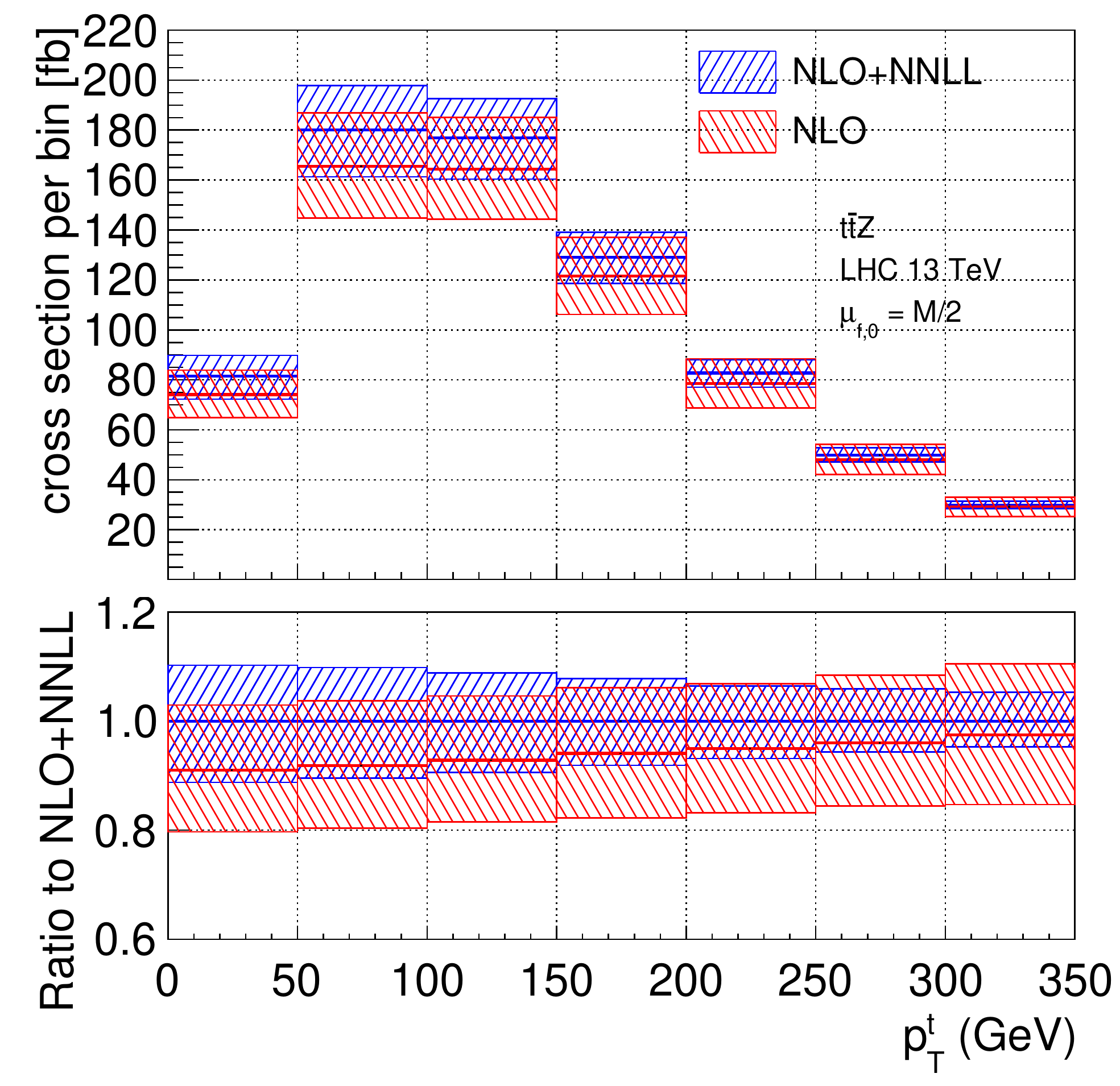} & \includegraphics[width=7cm]{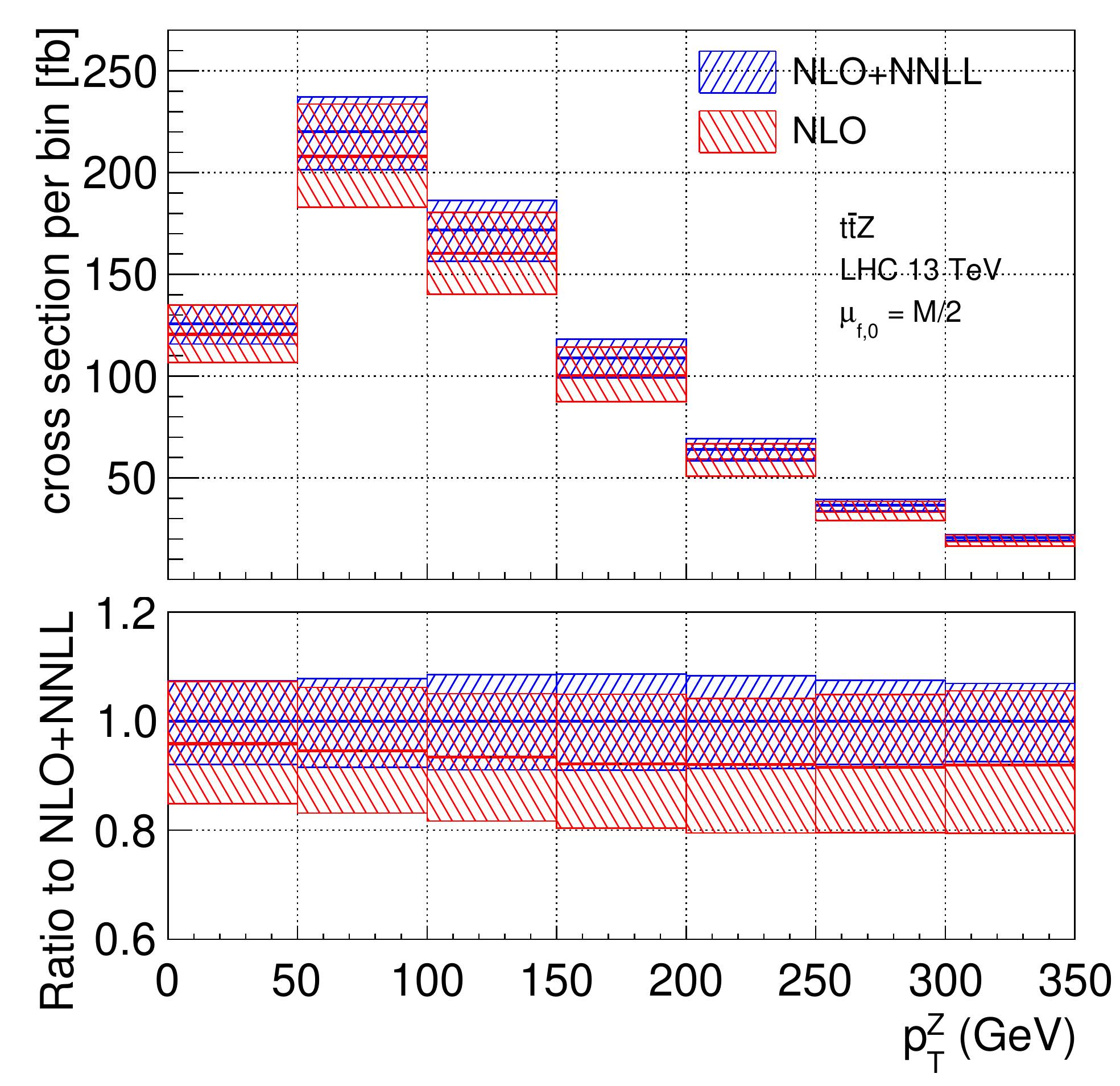} \\
		\end{tabular}
	\end{center}
	\caption{Differential distributions for $t \bar{t}Z$ production with $\mu_{f,0}=M/2$ at NLO+NNLL (blue band) compared to the NLO calculation (red band). The uncertainty bands are generated through scale variations of $\mu_f$, $\mu_s$ and $\mu_h$ as explained in \cite{Broggio:2017kzi}.
		\label{fig:NLOvsNNLLhalfMttZ}
	}
\end{figure}

In Figures \ref{fig:NLOvsNNLLhalfMttH} and \ref{fig:NLOvsNNLLhalfMttZ} we present four important differential distributions for the $t \bar{t} H$ and $t \bar{t} Z$ processes (similar plots are available for the $t \bar{t} W^\pm$ processes). In particular we compute the invariant mass distribution of the three heavy particles in the final state, the invariant mass of the $t \bar{t}$ system, the $p_T$ of the top quark and the $p_T$ of the heavy boson. We compare our best predictions to the fixed-order NLO calculations of the same observable.
The central values of the NLO+NNLL results are slightly larger than the central values of the NLO calculations in all bins. As expected, this effect is even more enhanced in the tails of the $M$ and $M_{t \bar{t}}$ distributions.
We also find that the scale uncertainty is nicely reduced after including the soft gluon resummation corrections. Moreover the NLO+NNLL error bands are usually contained in the upper part of the NLO uncertainty bands. This particular feature is common to all of the three processes.

\begin{table}[t]
	\begin{center}
		\def\arraystretch{1.3}
		\begin{tabular}{|c|c|c|c|}
			\hline  pert. order & process & PDF order & $\sigma$ [fb] \\
			\hline NLO & $t \bar{t} H$ & NLO & $474.8^{+47.2}_{-51.9} $ \\
			\hline NLO+NNLL & $t \bar{t} H$ & NNLO & $486.4^{+29.9}_{-24.5} $ \\
			\hline NLO & $t \bar{t} W^+$ & NLO & $356.3^{+43.7}_{-39.5} $ \\
			\hline  NLO+NNLL & $t \bar{t} W^+$ & NNLO & $341.0^{+23.1}_{-13.6} $ \\
			\hline NLO & $t \bar{t} W^-$ & NLO & $182.2^{+23.1}_{-20.4} $ \\
			\hline  NLO+NNLL & $t \bar{t} W^-$ & NNLO & $177.1^{+12.0}_{-6.9} $ \\
			\hline NLO & $t \bar{t} Z$ & NLO & $728.3^{+93.8}_{-90.3} $ \\
			\hline  NLO+NNLL &$ t \bar{t} Z$ & NNLO & $777.8^{+61.3}_{-65.2} $ \\
			\hline 
		\end{tabular} 
		\caption{Total cross section for $t \bar{t} H$, $t \bar{t} W^{\pm}$ and $t \bar{t} Z$ production at the
			LHC with $\sqrt{s} = 13$~TeV and MMHT 2014 PDFs. The default value
			of the factorization scale is $\mu_{f,0}=M/2$, and
			the uncertainties are estimated through
			variations of this scale together with the resummation scales
			$\mu_s$ and $\mu_h$.
			\label{tab:ttZttW}}
	\end{center}
\end{table}

\section{Conclusions}

In this talk we present recent results on the resummation of soft gluon emission corrections for the production of a top-quark pair in association with a heavy boson ($H,Z,W$) at the LHC. The resummation was implemented to NNLL accuracy in the partonic threshold limit. Numerical predictions for the cross sections and the differential distributions were obtained by means of an in-house parton level Monte Carlo program. 
We found that the impact on the central value of the cross sections is moderate for the particular choice of the factorization scale made in this work.
In addition, the residual perturbative uncertainty of the NLO+NNLL calculations (obtained by varying the hard, soft and factorization scales) is smaller than the NLO scale uncertainty.

We plan to extend our work in different directions. We are currently working on combining the most up-to-date QCD predictions (at NLO+NNLL accuracy) together with the EW corrections for this class of processes. The goal is to obtain the most complete determinations for the total cross sections and some important differential distributions in the SM.
Another interesting extension in the spirit of \cite{Broggio:2014yca} would be the inclusion of decay products of the heavy final state particles in the narrow width approximation.
In this way it would be possible to put kinematic cuts directly on the momenta of the detected particles.

\section*{Acknowledgments}
I would like to thank my collaborators A.~Ferroglia, G.~Ossola, B.~Pecjak, R.~Sameshima, A.~Signer and L.~Yang for working with me on these projects.

\end{document}